*Original Article*

# Securing Data Platforms: Strategic Masking Techniques for Privacy and Security for B2B Enterprise Data

Mandar Khoje

*Independent Researcher, Dublin, California, USA.*

*Corresponding Author : mandar.khoje@gmail.com*

*Abstract* - In today's digital age, the imperative to protect data privacy and security is a paramount concern, especially for business-to-business (B2B) enterprises that handle sensitive information. These enterprises are increasingly constructing data platforms, which are integrated suites of technology solutions architected for the efficient management, processing, storage, and data analysis. It has become critical to design these data platforms with mechanisms that inherently support data privacy and security, particularly as they encounter the added complexity of safeguarding unstructured data types such as log files and text documents. Within this context, data masking stands out as a vital feature of data platform architecture. It proactively conceals sensitive elements, ensuring data privacy while preserving the information's value for business operations and analytics. This protective measure entails a strategic two-fold process: firstly, accurately pinpointing the sensitive data that necessitates concealment, and secondly, applying sophisticated methods to disguise that data effectively within the data platform infrastructure. This research delves into the nuances of embedding advanced data masking techniques within the very fabric of data platforms and an in-depth exploration of how enterprises can adopt a comprehensive approach toward effective data masking implementation by exploring different identification and anonymization techniques.

## 1. Introduction

B2B enterprises provide value to other companies, many of which ingest customers' data into their data platforms to derive insights, analytics, and value for the customer. These sophisticated platforms encompass different types of data stores, ensuring the data is not only stored in various formats, such as raw text files, parquet files, or other structured and unstructured formats but is also accessible for further processing. Within these platforms, the data is then leveraged by downstream services to build data products, generate analytic insights, and, increasingly, train machine learning models. Given the sensitive nature of this data, which may include personal information or proprietary business details — for instance, analytics on logs of customer services might contain IP addresses, customer IDs, names, emails, and other forms of personally identifiable information (PII) — it poses significant risks if compromised. The repercussions of such a breach could lead to identity theft, financial fraud, or severe reputational damage for both the data-collecting enterprise and its customers. Existing research primarily focuses on dynamic data masking, a process of obscuring data at the time of retrieval or reading. However, as highlighted in this research, it is also essential to consider masking data directly within the data store, a point further elaborated in the literature review section.

Hence, B2B enterprises are tasked with implementing robust data protection measures within their data platforms to safeguard the privacy and security of their customers' information. In the context of B2B relationships and data management, the entities involved are typically classified as either "data controllers" or "data processors," particularly within the framework of legislation like the General Data Protection Regulation (GDPR). [1] The data controller is the entity that determines the purposes and means of the personal data processing.

In a B2B relationship, this would be the company that owns or has primary authority over the data. Conversely, the data processor is an entity that processes personal data on behalf of the controller, following the data controller's instructions. In these B2B engagements, the data processor could be a company performing services on the data — such as analytics or cloud services — within the data platform as per the controller's directives. It is, therefore, crucial to ensure data security as it flows within and between these various entities. This paper will specifically explore how integrating data masking techniques into data platforms can serve as a powerful strategy to enhance the protection of sensitive information in these complex B2B ecosystems.

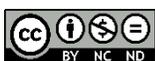

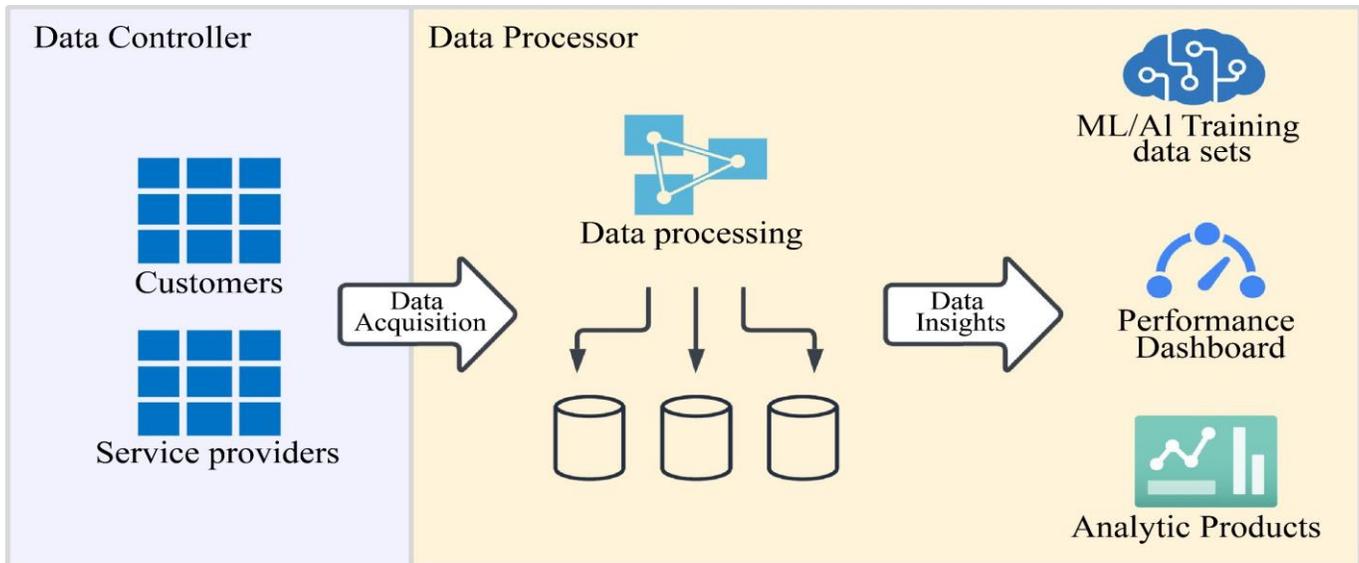

**Fig. 1 B2B enterprise data architecture**

## 2. Literature Review

Despite the widespread use of sophisticated data platforms, most existing research predominantly focuses on the masking of structured data, often employing dynamic masking techniques. This leaves a notable research gap in handling sensitive unstructured data, where dynamic masking may not be feasible due to performance constraints or other considerations.

The following research paper, "A Study on Dynamic Data Masking with its Trends and Implications" [2], describes how traditional static data masking solutions, which store obfuscated values in databases, have been limited to development and QA environments. It introduces a new approach for production environments: Dynamic Data Masking. This method enhances application security by applying various actions like masking, hiding, or blocking data without altering applications or physical databases. This approach reduces data exposure risks and is considered the best practice for securing production databases. Although, with the rise of ML and data analytics, more and more engineers need access to production data in order to build meaningful data products or to train ML models. This exposes the risk of sensitive data being handled by engineers or other employees. This is more evident in smaller enterprises that do not have different non-production environments. Therefore, his paper proposes a more proactive strategy. It suggests the upfront identification and masking of sensitive information as it is written to data stores, thus preventing inadvertent access to such data during processing. Furthermore, it introduces varied techniques for recognizing and masking sensitive data, offering the flexibility to treat different types of Personally Identifiable Information (PII) in unique ways. For instance, Personally Identifiable Information (PII) such as Social Security Numbers (SSN) is extremely sensitive and should generally never be unmasked, warranting full redaction. In contrast, an email address might be masked by only obfuscating the prefix, allowing for the possibility of unmasking later if required. This approach differs significantly from the handling of SSNs, which should remain perpetually masked to ensure maximum security. This approach represents a shift from traditional methods, focusing on enhancing security measures right from the data entry stage.

## 3. Methodology

Data masking, also known as data obfuscation or data anonymization, is a process used to protect sensitive information from unauthorized access by obscuring it to maintain the data's usability for legitimate purposes. [3] The purpose of data masking is to create a sanitized version of the data that can be used for purposes such as user training, software testing, or analytics without exposing the actual sensitive data.

Masking of data can be thought of in two steps. First, identifying sensitive information in the raw data that needs to be protected, and then applying a technique to transform or conceal that data while still preserving its usefulness.

### 2.1. Step 1: Identification of Sensitive Information
The first step involves identifying the sensitive information within the raw data. Several approaches can be leveraged to perform this step.

#### 2.1.1. Regex based Approach
Regular expressions [4] serve as a powerful tool for detecting patterns and specific formats indicative of sensitive information, such as social security and credit card numbers.

They are quite effective when dealing with data that adheres to known patterns; however, their efficacy diminishes when faced with variations or emerging forms of sensitive information. To mitigate such limitations, it is advisable to implement a tiered system of regular expressions, each with a designated confidence level based on the likelihood of the text being personally identifiable information (PII). For instance, in the context of obfuscating credit card numbers within a text corpus, a basic regular expression like ^\d{16}$ can be employed to identify sequences of 16 digits. This rudimentary method, though, is prone to a high rate of false positives, as not all 16-digit strings represent credit card numbers, rendering it a low-confidence approach.

Consider a more refined regular expression:
^\(\d{3}\) \d{3}-\d{4}$
The components of this expression are:
^ asserts the beginning of the string.
\( and \) precisely match the literal parentheses.
\d{3} requires a sequence of exactly three digits.
Spaces are used to separate different segments of the number.
\d{4} necessitates a sequence of exactly four digits at the end.
$ asserts the end of the string.

Designed to match phone numbers formatted as "(xxx) xxx xxxx", this expression demonstrates increased specificity and, thus, a higher confidence level. Nevertheless, it may yield false negatives by not accounting for alternative phone number formats, for example, phone numbers for different countries.

This underscores the need for a balanced and adaptable regular expression strategy that minimizes the chances of both false positives and negatives, tailored to the formatting conventions of the data set in question.

Many organizations may not have the necessary resources or infrastructure to leverage more advanced techniques, which are discussed below. For these organizations, as long as they are aware of the limitations, this can provide a good starting point to mask simple PII types, for example, email addresses, which mostly have a fixed format.

### 2.1.2. Bloom Filters

A Bloom filter [5] is a sophisticated probabilistic data structure that finds widespread application in the realms of computer science and information retrieval. It is engineered to ascertain if an element is possibly a member of a set, allowing for a certain degree of error margin in the form of false positives.

This error-tolerant feature means that while it can confidently indicate that an element is not part of a set, it cannot guarantee the presence of an element with absolute certainty; rather, it can only suggest its potential inclusion.

The strength of a Bloom filter lies in its spatial efficiency, making it highly advantageous for situations where storage is a premium. It provides a time-efficient means of querying membership status, which, for practical purposes, translates into swift determinations about an element's potential inclusion in a set without the need for exhaustive searches.

In the context of safeguarding Personally Identifiable Information (PII), such as individual names, one could leverage a Bloom filter by initializing it with a comprehensive database of names. This initialized filter can then serve as a checkpoint to evaluate whether a given string—potentially a name—is likely a part of this set. When a string is checked against the Bloom filter, if the outcome is negative, one can be assured that the name is not in the database. However, if the result is positive, there remains the slim possibility that it is a false positive.

This technique, while effective, presents several performance-related considerations. The dimensions of the Bloom filter—namely, its size and the selection of hash functions—directly impact its accuracy and efficiency. A smaller Bloom filter, although space-efficient, may yield higher false positives, whereas a larger one might be prohibitive regarding memory consumption. Similarly, the time complexity of the membership checks, although typically fast, can vary depending on these parameters and the implementation of the hash functions.

Moreover, this approach has inherent limitations in dealing with anomalous data, such as misspelt names or names absent from the initial database used to create the filter. These instances will invariably lead to names not being identified and, consequently, not masked. It is crucial, therefore, to maintain an extensive and up-to-date name database to minimize such oversights.

In summary, while a Bloom filter is an elegant and efficient solution for potential membership querying and PII masking, it necessitates careful consideration of its configuration and limitations. Optimal utilization of a Bloom filter for PII protection hinges on a balance between space complexity and the probability of false positives while being cognizant of the nuances inherent in the data it is employed to protect.

### 2.1.3. ML Models

Employing machine learning (ML) techniques for the purpose of data masking represents a significant advancement over traditional methods, offering enhanced accuracy and higher confidence in the identification and protection of Personally Identifiable Information (PII). By developing and fine-tuning machine learning models that are meticulously trained to recognize various forms of PII, organizations can achieve a more sophisticated level of data privacy.

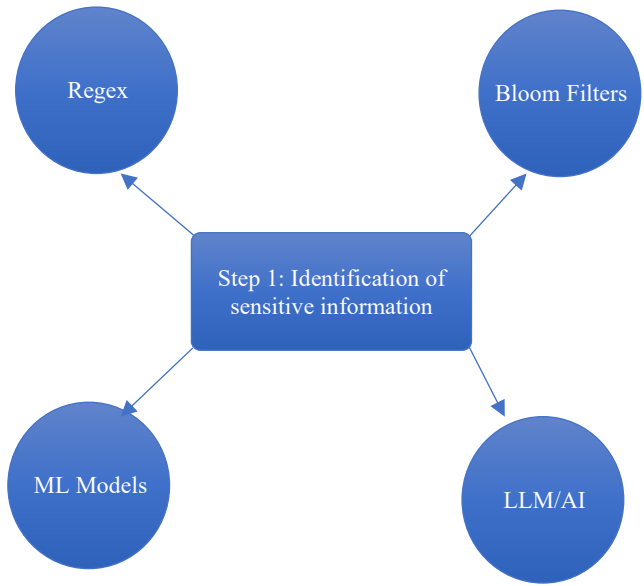

**Fig. 2 Step 1: Identification of sensitive information**

The process of constructing such a model necessitates a dataset that has been previously annotated with examples of PII. This dataset serves as the foundation upon which the model learns to discern between sensitive and non-sensitive information. The machine learning algorithm can uncover complex patterns and nuances that may elude simpler rule-based systems like regular expressions through exposure to this diverse set of training examples.

One of the key strengths of machine learning models in this application is their capacity for high precision in PII identification. As the model iteratively processes the annotated data, it refines its ability to accurately classify new, unseen instances of information as PII or non-PII. This iterative learning process is central to the model's development, enabling it to understand the intricacies and variations present in human names, addresses, social security numbers, and other forms of personal data.

Moreover, machine learning models are inherently adaptable. As new forms of PII emerge or patterns evolve over time, the model can be retrained or fine-tuned with updated data to maintain its efficacy. This adaptability is critical in the ever-changing landscape of data privacy and security, ensuring that the model remains robust against novel challenges.

It is important to recognize, however, that the success of a machine learning approach for PII masking heavily depends on the quality and representatives of the training data. A well-annotated dataset that captures a wide array of PII scenarios is essential for the model to learn effectively. Additionally, care must be taken to ensure that the model does not simply memorize the training data but generalizes its understanding to identify PII in varied contexts accurately.

One research shows that fine-tuned machine learning models can achieve higher accuracy and performance compared to other automated ways of scanning PII. [6]

In conclusion, machine learning models offer a dynamic and effective solution for data masking, capable of achieving superior levels of accuracy and adaptability when it comes to the identification of PII. When appropriately trained and maintained, these models stand as a cutting-edge tool in the arsenal of data privacy and protection strategies.

*2.1.4. LLMs/AI*
The advent of Large Language Models (LLMs) [7], such as those developed by OpenAI, marks a transformative shift in the approach to data masking, particularly for the identification and redaction of Personally Identifiable Information (PII). These advanced models, pre-trained on extensive corpuses of text, bring a level of contextual understanding and flexibility previously unattainable with methods such as regular expressions or Bloom filters.

Unlike specialized models that require a substantial investment in data collection, annotation, and training, LLMs can be harnessed directly through carefully crafted prompts. By instructing an LLM with prompts designed to identify and mask PII, users can leverage the model's vast knowledge base and its nuanced understanding of language.

This approach benefits from the LLM's ability to interpret the context surrounding PII, discerning between sensitive and benign data with high precision.

For example, when provided with a prompt that asks the LLM to "identify and replace sensitive information in the following text with asterisks," the model can process the request and execute the task with remarkable accuracy. This is due to its underlying training, which enables it to recognize patterns of language that characterize PII.

The LLM's performance in this regard surpasses regular expressions, which are rigid and unable to account for the myriad ways in which PII can be presented in text. Furthermore, it addresses the limitations of Bloom filters, which cannot handle misspellings or variations not present in their initial configuration. With their capacity for understanding and learning from context, LLMs can effectively manage these challenges.

Moreover, the utilization of an LLM for data masking does not suffer from the binary limitations of a Bloom filter's possible false positives. Instead, it offers a probabilistic and nuanced assessment, providing confidence scores that allow for a more informed decision-making process regarding what constitutes PII in each unique instance.

Regarding scalability and ease of use, LLMs also present a distinct advantage. There is no need for the extensive computational resources often associated with training and running specialized machine learning models. Instead, LLMs can be accessed via APIs or cloud services, offering high accessibility and quick integration into existing data protection workflows. It is important to note that the LLM services being used for PII recognition may or may not use that data to train its model. For some organizations, this can be an impediment as it poses a security risk.

In conclusion, using Large Language Models for PII masking represents a significant leap forward in both the efficacy and efficiency of data protection measures. By utilizing these models through intelligent prompting, organizations can achieve a more accurate, context-aware, and cost-effective means of safeguarding sensitive information.

*2.2. Step 2: Masking of Sensitive Information*
Once the PII spans are identified in the raw text, this additional information can be sent to the data masking component, which applies appropriate techniques to mask the identified PII while preserving the overall structure and integrity of the data. There are different approaches that could be taken to achieve this.

*2.2.1. Redaction*
Masking using redaction typically refers to the process of removing or obscuring sensitive information from a document or a dataset. This method is commonly used in legal documents, government records, and corporate communications, where sensitive information must be protected from disclosure. Unlike hashing or encryption, redacted information is permanently removed or concealed and cannot be restored.

Once PII is identified, this information is either blacked out, removed, or replaced with placeholder text such as "[REDACTED]". This process can be done manually or automatically with software tools designed for redaction.

It is important to ensure the redaction cannot be reversed, especially if the document is in a digital format.

Secure Disposal: If there were any copies or versions of the document that contained the sensitive information before redaction, they need to be securely disposed of.

Redaction is a common technique used in data masking to hide sensitive information by replacing it with placeholder values. For example, in the string "My phone number is 111-111-1111", using redaction for phone numbers, the string would be "My phone number is <PHONE_NUMBER>" if our placeholder value for PII of type phone numbers is <PHONE_NUMBER>.

*2.2.2. Anonymize*
Another approach is to anonymize the PII by replacing it with generic or randomized values. An example, "My phone number is 111-111-1111", an anonymized string would be "My phone number is 434-232-3453". This number is a randomly generated number.

*2.2.3. Encryption*
Masking using encryption is a data protection method that transforms original data into another form, or ciphertext, so only authorized users can reverse the transformation. Unlike hashing, a one-way process, encryption is designed to be reversible using the appropriate decryption key.

There are two primary types of encryption algorithms used for masking:

Symmetric Encryption [8]: In symmetric encryption, the same key is used for both encryption and decryption. AES (Advanced Encryption Standard), DES (Data Encryption Standard), and RC4 are some of the examples.

Asymmetric Encryption [9]: This uses a pair of keys: a public key for encryption and a private key for decryption. RSA, ECC (Elliptic Curve Cryptography), and PGP (Pretty Good Privacy) are common asymmetric encryption algorithms.

Here is an overview of the process for symmetric encryption:
Key Generation: A secure key is generated. For strong encryption, this key should be of sufficient length (e.g., AES-256 uses a 256-bit key).

Encryption: Using the encryption algorithm and the key, the original data (plaintext) is encrypted and turned into ciphertext.

Storage/Transmission: The encrypted data can be safely stored or transmitted. It is unreadable to anyone who does not have the key.

Decryption: The recipient, who must have the decryption key, can then decrypt the ciphertext back into the readable original data (plaintext).

Key Management: The encryption and decryption keys must be securely managed. If a symmetric key is used, it needs to be securely transferred to the recipient.

The use of encryption allows for data to be securely masked while in transit or at rest and also be fully recoverable when needed. This is particularly useful for data that needs to be processed or read by authorized individuals or systems.

*2.2.4. Hashing*

Masking using hashing [10] is a technique to obscure original data (such as personally identifiable information or PII) by using a hash function. This technique is often used to protect sensitive data while maintaining a level of utility. The hash function takes input data and produces a fixed-size string of characters, typically appearing randomly. This process is one-way, meaning reversing the hash back to the original data should not be feasible.

Here is an overview of the process:
Selection of Hash Function: Choose a strong cryptographic hash function like SHA-256. The hash function should be a one-way function that produces a unique output for each unique input.

Salting (optional but recommended): Before hashing, you can add a "salt" to the data, a random value used to ensure that the same input does not always result in the same hash.

This helps protect against rainbow table attacks. [11]
Hashing: Run the data through the hash function. The function will produce a fixed-length string of characters.

Storing Hashes: Store the hash values in place of the original data. If salts are used, they must also be stored to allow for future verification or comparisons.

Verification: To verify a piece of data against the hashed value, you hash the new piece of data with the same hash function (and salt, if used) and compare the result to the stored hash. If they match, the data is verified.

It is important to note that while hashing can be an effective way to mask data, it is not always the best approach for all scenarios. Hash functions are deterministic, meaning the same input will always produce the same output.

This can be problematic if the set of possible inputs is small or if the inputs are guessable, as attackers could potentially use a brute force approach to generate hashes of possible inputs and match them to the stored hashes. That is why salting is an important part of the process when handling sensitive information.

Additionally, hashing is unsuitable for data that needs to be decrypted back to its original form since it is a one-way function. For such cases, encryption with a key would be a more appropriate method, as encrypted data can be decrypted with the correct key.

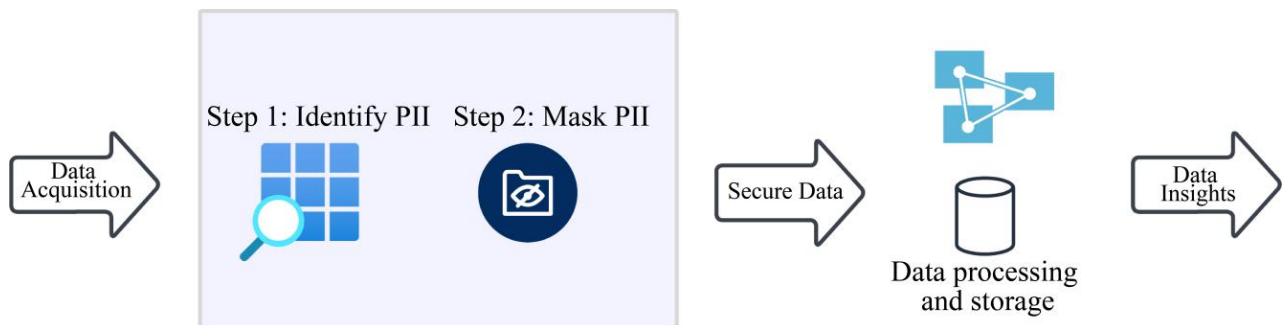

**Fig. 3 PII detection and masking**

*2.2.5. Custom*

Specifying a custom function to mask data is a practical approach, particularly for cases where certain data elements must be preserved to maintain the informational context. At the same time, other parts require obfuscation for privacy reasons.

In the context of email addresses, this means obscuring the local part (the section before the "@") while keeping the domain part visible to indicate the service provider or organizational domain, which can be useful for demographic or categorical analyses.

Given the string "My email address is johndoe@example.com", a custom masking function can be designed to replace "johndoe" with a placeholder such as "xxxxxxx" while leaving the domain "@example.com" intact. The resulting masked string would be "My email address is xxxxxxx@example.com". This method retains the structural integrity of the email address, indicating that it is indeed an email while protecting the user's identity.

Here is a step-by-step explanation of how such a custom function works:

Segmentation: The function splits it into two parts at the "@" symbol: the local part requiring masking and the domain part remaining unmasked.

Replacement: The function replaces the local part with a series of "x" characters or another placeholder symbol. The length of this series can match the length of the original local part or be a fixed length for consistency.

Concatenation: The masked local part is then concatenated with the original domain part to produce the final anonymized email address.

Integration: The new masked string is reinserted into the original text, replacing the original email address.

When designing such a function, it is important to ensure that the masking process is irreversible, meaning there should be no practical method to reconstruct the original information from the masked version. This is crucial for compliance with data protection standards and for maintaining the trust of individuals whose data is being handled.

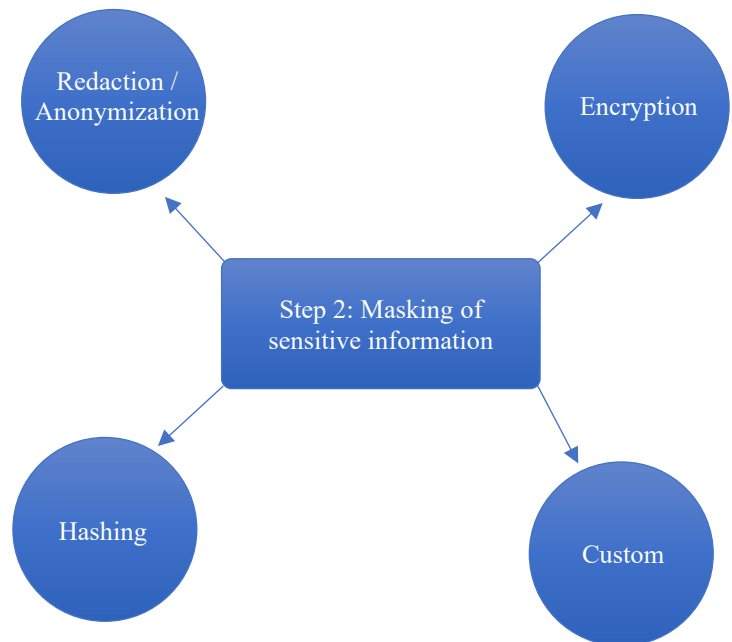

**Fig. 4 Step 2 : Masking of sensitive information**

## 3. Limitations

The limitations of this research are pronounced within the context of the evolving architecture of data platforms and the multifaceted nature of data security, particularly with unstructured data. While this paper outlines an array of techniques for detecting and masking PII within data platforms, it does not claim to be exhaustive. The swiftly changing technology landscape and the advent of new cybersecurity threats present ongoing challenges that may not be fully encompassed by the methodologies currently detailed.

Moreover, the research touches on but does not thoroughly investigate the performance implications of data masking within these platforms. The fine line between maintaining robust data protection and ensuring system efficiency is a delicate one that merits further exploration. The precision of PII detection methods is also a critical factor; however, the potential for false positives or negatives, which could impact the utility of the data and the integrity of privacy safeguards, is recognized but not conclusively addressed in this study.

The paper touches on the technical, financial, and resource considerations of implementing the recommended data masking strategies in data platforms but does not delve into these aspects comprehensively. Additionally, while legal and regulatory complexities in data privacy are acknowledged, a thorough engagement with these topics across various jurisdictions is outside the paper's current scope.

The inherent challenge of preserving the utility of masked data for analytical and machine-learning purposes is recognized, yet it is not exhaustively treated. As such, the applicability of the recommendations to all B2B enterprises may be limited due to variations in industry practices, types of data, and the scale of operations.

Finally, the paper proposes recommendations for integrating data masking practices within data platforms but does not extend to their practical implementation and validation in diverse real-world environments. This underlines a broader acknowledgement that this study is but an initial step towards comprehensively integrating data protection strategies within data platforms.

It underscores the necessity for continuous adaptation and enhancement in data security measures, ensuring they evolve in tandem with technological progress and emerging threats. This ongoing evolution is essential to maintain the relevance and efficacy of privacy measures in an ever-advancing digital world.

## 4. Conclusion and Future Research Directions

In conclusion, the imperative for B2B enterprises to protect sensitive customer data within their data platforms is a regulatory and ethical necessity and a strategic business requirement. The architecture and functionality of these platforms must inherently address the complex challenges of data privacy and security, especially when handling personal and proprietary business information. This research has presented a comprehensive methodology for identifying and masking personally identifiable information (PII) and sensitive data within these multifaceted platforms, particularly when dealing with unstructured data formats.

The initial phase involves the meticulous detection of sensitive elements within the data platform, leveraging innovative approaches such as regular expressions, Bloom filters, machine learning algorithms, and the utilization of sophisticated large language models. Subsequently, the data masking process is operationalized through a plethora of techniques, including encryption, hashing, redaction, and the deployment of bespoke functions that align with the enterprises' unique demands and operational contexts.

Effectively integrating these data masking strategies into data platforms enables B2B companies to balance the exploitation of vast data reservoirs with the imperative of privacy and security. This equilibrium is crucial for safeguarding against data breaches and their legal and reputational consequences and cultivating trust with customers and partners through demonstrable commitments to information security.

Moreover, as data remains a pivotal asset for innovation and business growth—underscored by the burgeoning field of data-centric services and machine learning models—the indispensability of fortified data protection mechanisms becomes increasingly pronounced. B2B enterprises that proactively incorporate sophisticated data masking solutions into their data platforms are poised to secure a competitive advantage, creating an ecosystem where data is harnessed responsibly and productively.

Finally, while this paper charts a course toward robust data protection within data platforms through effective masking strategies, it simultaneously serves as a clarion call for ongoing vigilance and evolution. As technological landscapes shift and new threats materialize, it is incumbent upon enterprises to continually refine their data protection frameworks. The quest for comprehensive data security is a perpetual endeavor, necessitating sustained investment in both technological solutions and a corporate ethos that elevates the importance of data privacy. Through this unyielding commitment, businesses will navigate the digital epoch of the future with confidence and integrity.